\documentclass[runningheads,a4paper]{llncs}
\usepackage[table,xcdraw]{xcolor}
\usepackage{amssymb}
\usepackage{graphicx}
\usepackage{xcolor}

\newcommand{\keywords}[1]{\par\addvspace\baselineskip
\noindent\keywordname\enspace\ignorespaces#1}

\begin{document}

\mainmatter

\title{AI for Calcium Scoring}

\titlerunning{AI for Calcium Scoring}

\author{Sanne G.M. van Velzen\textsuperscript{1,2}\and Nils Hampe\textsuperscript{1,2}\and Bob D. de Vos\textsuperscript{1,2}\and Ivana I\v{s}gum\textsuperscript{1,2,3}}
\institute{\textsuperscript{1}Department of Biomedical Engineering and Physics, Amsterdam University Medical Centers - location AMC, University of Amsterdam, the Netherlands\\
\textsuperscript{2} Amsterdam Cardiovascular Sciences, Amsterdam University Medical Centers, the Netherlands\\
\textsuperscript{3}Department of Radiology and Nuclear Medicine, Amsterdam University Medical Centers - location AMC, University of Amsterdam, the Netherlands}

\authorrunning{Sanne G.M. van Velzen\and Nils Hampe\and Bob D. de Vos\and Ivana I\v{s}gum}

\maketitle

\begin{abstract}
Calcium scoring, a process in which arterial calcifications are detected and quantified in CT, is valuable in estimating the risk of cardiovascular disease events. Especially when used to quantify the extent of calcification in the coronary arteries, it is a strong and independent predictor of coronary heart disease events. Advances in artificial intelligence (AI)-based image analysis have produced a multitude of automatic calcium scoring methods. While most early methods closely follow standard calcium scoring accepted in clinic, recent approaches extend this procedure to enable faster or more reproducible calcium scoring. This chapter provides an introduction to AI for calcium scoring, and an overview of the developed methods and their applications. We conclude with a discussion on AI methods in calcium scoring and propose potential directions for future research.

\end{abstract}

\section{Introduction}

Cardiovascular disease (CVD) accounts for approximately one-third of all global deaths \cite{roth2018global}. With obesity being a global health concern, worldwide rates of diabetes mellitus on the rise, and an increasing number of smokers in developing countries, the need for early detection of CVD risk is increasingly apparent.

Calcium scoring, a process in which arterial calcifications are detected and quantified in CT, has been shown to be valuable in prediction of cardiovascular events \cite{greenland2018coronary}. For instance, calcifications in the aortic arch and carotid arteries have been associated with an elevated risk of stroke \cite{iribarren2000calcification,nandalur2006carotid}. The amount of coronary artery calcification (CAC) is a strong and independent predictor of coronary heart disease (CHD) events\cite{hecht2015coronary,budoff2009coronary} and therefore, calcium scoring is most often applied to the coronary arteries. 

Calcium scoring is standardly performed in CT scans without contrast enhancement, where lesions are defined as a high density area of $\geq$ 130 Hounsfield Units (HU) in an artery\cite{agatston}. Calcifications are standardly quantified into Agatston, volume and mass scores \cite{rumberger2003rosetta}. The volume score represents the volume of calcified lesions and it only takes the voxel size into account, while the calcium mass considers both the volume and the density of the lesions. Both scores are used for quantification of calcifications in any artery, while the Agatston score is applied to the coronary arteries only. The Agatston score is a slice based quantification method that combines the area of a lesion with an ordinal weight term based on the maximum density of the lesion in each slice ( $<200$HU: 1; $<300$HU: 2; $<400$HU: 3; $\geq400$HU: 4). The total Agatston score is obtained by summing the 2D lesion scores over all slices. Based on the total Agatston score a patient can be assigned to a risk category that directly relates to a 10 year risk of CVD events \cite{agatston,budoff2018ten}. Commonly used risk categories are listed in Table \ref{tab:RC}.

\begin{table}[]
\centering
\label{tab:RC}
\caption{Commonly used risk categories based on Agatston scores determind in the coronary arteries.}
\begin{tabular}{lllc}
\hline
\rowcolor[HTML]{FFFFFF} 
\multicolumn{2}{l}{\cellcolor[HTML]{FFFFFF}\textbf{Agatston score}} & \textbf{} & \multicolumn{1}{c}{\cellcolor[HTML]{FFFFFF}\textbf{Risk}} \\ \hline
\rowcolor[HTML]{FFFFFF} 
                         & 0                                        &           & very low                                                               \\
\rowcolor[HTML]{E9EAFC} 
                         & 1-10                                     &           & low                                                                    \\
\rowcolor[HTML]{CBCEFB} 
                         & 11-100                                   &           & intermediate                                                           \\
\rowcolor[HTML]{B1B2EF} 
                         & 100-400                                  &           & high                                                                   \\
\rowcolor[HTML]{9698ED} 
                         & \textgreater{}400                        &           & very high                                                              \\ \hline
\end{tabular}
\end{table}

\subsection{Standard calcium scoring}

Routinely, only coronary calcium scoring is performed. In clinic, this is done with commercially available software in a semi-automatic manner. The interactive software projects an overlay showing all the areas above 130 HU onto a CT slice. Subsequently, the areas in the coronary arteries representing CAC are manually identified by the expert with a mouse click. The software then automatically identifies the whole lesion and labels it according to the artery label provided by the expert. This process is repeated for each CAC lesion and each CT slice. Thereafter, the identified lesions are quantified into Agatston score, volume and mass. The scores are calculated per artery and over all arteries. 

\begin{figure}[h]
    \centering
    \includegraphics[width=\textwidth, trim={0cm 28cm 0cm 1cm}, clip]{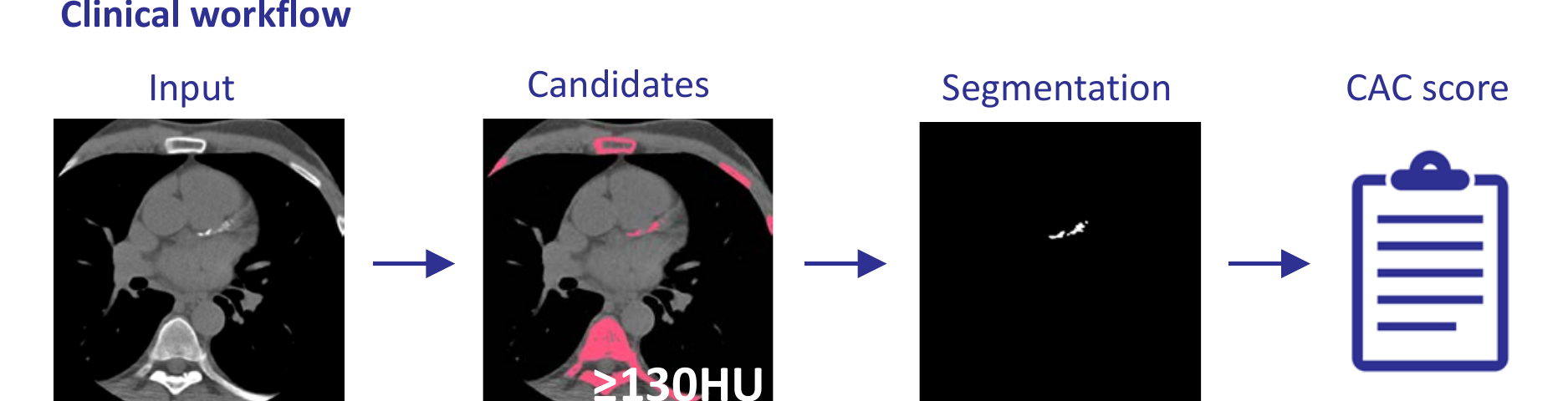}
    \caption{Clinical workflow for CAC quantification: Lesion candidates are extracted based on their image intensity values and thereafter, the arterial calcifications of interest are identified and quantified.}
    \label{fig:clinical_workflow}
\end{figure}

\subsection{AI for calcium scoring}
Although manual calcium scoring is not considered a difficult task for experts, it can be tedious and time consuming, especially when large numbers of scans are involved or the image quality is compromised. Therefore, a multitude of artificial intelligence (AI) methods have been developed to automate calcium scoring \cite{hampe2019machine,litjens2019state}. 

Due to large variability in the manifestation of calcified plaque and its look-alikes, designing rule-based algorithms for identifying calcified plaque is far from trivial. Instead, state-of-the art automatic systems utilize machine learning to learn the intuitive rules experts apply when scoring a scan. Calcium scoring is often posed as a classification problem where the machine learning task is to separate arterial calcifications from their lookalikes. In conventional machine learning, classification is performed based on the information describing the potential calcifications (features) that is relevant for the task at hand. In case of calcium scoring the potential calcifications typically consist of candidate lesions above 130 HU in the scan, that need to be classified into calcifications or other lesions. The features used in automatic calcium scoring typically describe lesion size, shape, texture and location in the scan. For calcium scoring it is feasible to obtain labels for the training data, i.e. for each candidate lesion a label can be obtained that indicates whether the lesion is a true arterial calcification or not. Hence, the training of calcium scoring methods is supervised: the classifier learns to perform the classification based on labeled example data. When training is finished, the free parameters are frozen and the algorithm can be used to classify previously unseen samples. 

While extensive research throughout multiple decades has produced a wide spectrum of machine learning algorithms, hardware and algorithm developments over the last years evoked the creation of an entirely new AI category, termed deep learning. The most distinct difference between deep learning and the conventional machine learning algorithms lies in their ability to learn directly from the data, circumventing the need for designing features. This shifts the focus from carefully designing specific features that incorporate expert knowledge about the task at hand to engineering the network architecture and the optimization process. Furthermore, ability of deep learning approaches to learn optimal features simultaneously with the classification can result in substantially improved performance. Like in calcium scoring using conventional machine learning, deep learning calcium scoring methods predominantly utilize supervised approaches. 

We discuss machine learning and deep learning methods for calcium scoring that follow standard identification of calcified lesions in clinic in Section 2. Later methods advance coronary calcium scoring to achieve faster and more reproducible measurements. These methods are discussed in Section 3. Finally, Section 4 provides a discussion on the current state of the art and potential future research directions. 

\section{Automatic coronary calcium scoring}
To date, most automatic coronary calcium scoring methods are closely related to the expert manual calcium scoring procedure using commercial software solutions: Lesion candidates are first extracted based on their image intensity values and thereafter, the calcifications of interest are identified (Fig. \ref{fig:clinical_workflow}). Currently, most automatic CAC scoring methods focus on automating the manual lesion identification with a machine learning classifier or deep learning architecture. 

In clinical practice, only coronary calcium scoring is routinely performed and it is done in dedicated calcium scoring CT (CSCT) scans. However, other CT protocols that visualize the heart also allow CAC scoring. Because different CT protocols pose different challenges, we describe automatic CAC scoring per target CT protocol.

\subsection{Coronary calcium scoring in cardiac CT}
\label{sec:dedicated}
For quantification of CAC, a dedicated CSCT is performed without contrast enhancement, with electrocardiography (ECG)-triggering, 120 kVp and 3 mm slice thickness \cite{hecht20172016}. The dedicated protocol allows for optimal visualization of coronary calcifications and hence, accurate quantification of CAC. Absence of contrast agent in the arteries allows for good visualization of the calcified lesions. Moreover, the slice thickness of 3 mm minimizes image noise and therefore allows discrimination between small calcifications and noise voxels. Furthermore, using ECG, image acquisition is triggered to the end-diastolic phase, when the heart is most stable. Minimizing cardiac motion is particularly important for coronary calcium scoring due to its pronounced effect on visualization of the coronary arteries. 

Analogous to clinical calcium scoring, earliest AI methods for coronary calcium scoring were generally focused on application in CSCT scans, in which they computed a total CAC score. The typical automation pipeline of these early methods starts with defining the region of interest (ROI). In the defined ROI, the methods identify true calcified lesions among a set of candidates using a supervised machine learning classifier based on handcrafted features \cite{Isgum2007,kurkure_supervised_2010,qian2010lesion,brunner_toward_2010,shahzad_vessel_2013,wolterink2015automatic}.

Published methods mainly differ in strategies for defining the region of interest (ROI), the choice of machine learning classifier and the type of computed features. A combination of classifiers typically led to the best performance \cite{Isgum2007,kurkure_supervised_2010}. Isgum et al \cite{Isgum2007} used k-nearest neighbor (kNN) classifiers sequentially, where the result of the first classifier was used as input to the second classifier:
In the first stage, obvious positives and obvious negatives were discarded, and, subsequently, a new classifier was trained for the remaining, more difficult candidates \cite{Isgum2007}. Alternative to such a sequential combination, Kurkure et al. \cite{kurkure_supervised_2010} proposed to merge the output of an ensemble of multiple support vector machine (SVM) classifiers that were trained separately on different independently sampled subsets of the training data. This way a robust method was trained.

Next to the classifier, the computed features played an essential role in the automation pipeline. Features often describe shape, size, appearance and location of calcium lesions \cite{Isgum2007,kurkure_supervised_2010,qian2010lesion,brunner_toward_2010,shahzad_vessel_2013,wolterink2015automatic}. While shape, size and appearance features are described directly from the lesions using for instance volume, Gaussian filters and intensity value statistics, their spatial location requires contextual information from the scan.
For instance, I\v{s}gum et al.\cite{Isgum2007} described the position of lesions with respect to anatomical structures, i.e. the heart and the aorta. Kurkure et al. \cite{kurkure_supervised_2010} defined location features relative to a bounding box around the heart. The features were agnostic to the specific machine learning classifier and their descriptive power was crucial for algorithm performance.

Besides location information being important for distinguishing true coronary calcifications among candidates, it is useful for computing per artery scores. Therefore, machine learning methods were proposed that, unlike the aforementioned methods, used location information to infer in which coronary artery a lesion may reside in.
Brunner et al.\cite{brunner_toward_2010} used coordinates with respect to manually placed landmarks as features to assign all voxels in a CSCT scan to classes belonging to one of the three main coronary artery branches (LAD, LCX and RCA) or background with an SVM. By assigning lesion candidates to the predicted coronary artery, artery specific calcium scoring was performed. Other methods utilized an additional set of coronary CT angiography scan (CCTA) scans, in which an intravenously administered contrast agent enables visualization of the arteries \cite{shahzad_vessel_2013,wolterink2015automatic}. The CCTA scans were used to define coronary artery atlases, i.e. probabilistic maps of artery locations in typical heart anatomies. Registration of these atlases to CSCT images allowed for obtaining estimates of the coronary artery location and assigning a lesion to an artery. In addition to enabling the computation of artery specific calcium scores based on the CCTA atlases, Shahzad et al. \cite{shahzad_vessel_2013} also exploited artery locations in the form of an additional location feature relative to the arteries. Due to the differences in contrast between CCTA and CSCT, accurate registration is a very challenging task. Therefore, Wolterink et al. \cite{wolterink2015automatic} proposed to first transfer the CCTA atlases to CSCT atlas scans. The hereby obtained CSCT atlases were subsequently used to estimate artery locations. Moreover, the impact of potential registration errors on classification performance was mitigated by computing location features with respect to all coronary arteries instead of merely the closest one. Furthermore, the authors enabled an additional opportunity for a semi-automatic workflow, in which high uncertainty output of the trained classifier was presented to an expert for potential correction. By referring only lesion candidates with high classifier uncertainty, performance similar to a second observer was achieved, with relatively little manual expert input \cite{wolterink2015automatic}.

Recently, deep learning methods have been proposed for CAC scoring in CSCT. Martin et al. \cite{martin2020evaluation} segmented the territories in CSCT that belong to the three main coronary arteries with a segmentation convolutional neural network (CNN). Subsequently, only candidate lesions in the vicinity of the coronary arteries were classified using a neural network that combines spatial coordinate features with features from the segmentation CNN.
Van den Oever et al. \cite{oever2020deep} proposed an approach that employs a CNN to segment CAC by analyzing image slices in three orthogonal directions and thereafter, combines them to obtain the final segmentation. The evaluation showed that due to the absence of false negative predictions, the method could be used to exclude scans without CAC to relieve the workload of radiologist.

Although direct comparison of attained performances by the presented methods is generally desirable, it is often not possible due to differences in evaluation metrics and used datasets. Therefore, Wolterink et al. \cite{wolterink2016evaluation} organized the orCaScore challenge that provided a set of multi-center and multi-vendor training and test CSCT and CCTA scans and a standardized evaluation of the results. At the workshop that launched the challenge, the performance of five (semi-)automatic calcium scoring methods aimed at CSCT was assessed, including the methods of Shahzad et al. \cite{shahzad_vessel_2013} and Wolterink et al. \cite{wolterink2015automatic}. The authors found that, on a lesion level, some methods can identify CAC in CSCT with a sensitivity and positive predictive value close to that of expert observers. Nevertheless, all evaluated methods made common errors, typically at the coronary ostia, where the exact starting point of the coronary arteries is potentially ambiguous, and in distal segments of the coronary arteries, that are difficult to extract with most coronary artery tracking methods.

\subsection{Calcium scoring in non-contrast CT showing the heart}
\label{sec:chest}

\begin{figure}[h]
    \centering
    \includegraphics[width=\textwidth, trim={2cm 20cm 1cm 1cm}, clip]{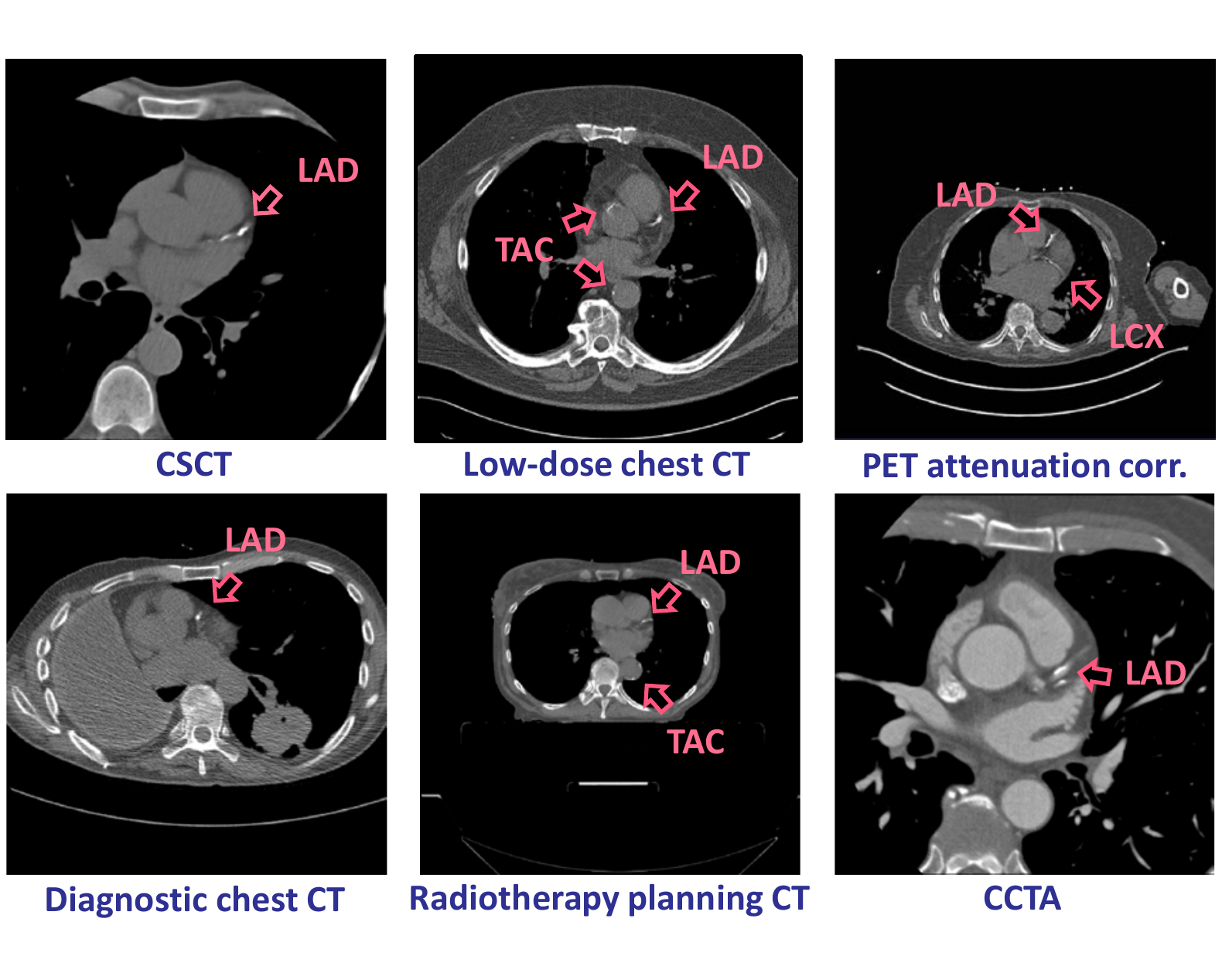}
    \caption{Examples of different CT scan protocols with pronounced differences in field of view, image resolution, cardiac motion, intravenous contrast and pathology. The arrows indicate calcified lesions in the left anterior descending (LAD) and left circumflex (LCX) artery, and thoracic aorta calcification (TAC).}
    \label{fig:examples}
\end{figure}

Over 7 million non-contrast chest CT scans are annually acquired for clinical care in the United States alone. This amounts to approximately 10 times more acquired chest CT scans than dedicated CSCT scans \cite{de2009projected}. Moreover, if lung cancer screening with chest CT would be considered, potentially another 7 to 10 million low-dose chest CTs per year would be acquired \cite{moyer2014screening}. Furthermore, non-contrast CT scans showing the chest are made for non-diagnostic purposes, e.g. radiotherapy treatment planning CT. All these scans visualize the heart and allow calcium scoring in the coronary arteries and in the aorta (Fig. \ref{fig:examples}). Given the clinical relevance of CAC scoring, the guidelines of the Society of Cardiovascular Computed Tomography and the Society of Thoracic Radiology \cite{hecht20172016} recommend to quantify and report CAC on \textit{all} such CT scans. 

However, a number of extra challenges arise that are not present in CSCT. Since these scans are made for a wide variety of clinical indications, the field of view is not focused on the heart but contains a larger volume, which necessarily translates to a lower in-plane resolution. Moreover, such CT scans are usually acquired without ECG-triggering, resulting in more pronounced cardiac motion artefacts impacting the appearance of calcifications in the coronary arteries.

Nevertheless, also in these scans calcium scoring has most frequently been applied to the coronary arteries. Because localization of the coronary arteries is not feasible due to lack of contrast enhancement and cardiac motion artefacts, machine learning methods for CAC scoring often simplify the analysis by defining a ROI or incorporate spatial information. For instance, González et al. \cite{gonzalez2016automated} defined a bounding box around the heart to determine the ROI using machine learning classifiers that classified the presence of the heart per slice in 3 orthogonal directions. Thereafter, a rule based approach was used to classify CAC lesions within the bounding box. Other methods did not specifically restrict the analysis to a certain ROI, but incorporated spatial information in a different manner. I\v{s}gum et al. \cite{isgum2012automatic} explicitly utilized the property that coronary calcifications appear at typical locations in the coronary artery tree. Through image registration of scans with reference segmentations of CAC, a spatial CAC probability map was created that was used to compute spatial features for candidate calcifications. Like in the methods for CAC detection in CSCT, features describing shape, size, appearance and location of the lesion were computed. These were used to detect CAC lesions with a combination of classifiers. Although the method was developed for low-dose chest CTs, the same method was also applied to attenuation correction CT scans for cardiac perfusion PET scans \cite{ivsgum2018automatic} and radiotherapy planning CT scans of breast cancer patients \cite{gernaat_automatic_2016}, both with good agreement with reference scores \cite{cohen1960coefficient}.

Owing to the typically large field of view, these scans also allow quantification of extracardiac arterial calcification in for instance the thoracic aorta, innominate artery or carotid arteries.
Accordingly, several automatic methods for extracardiac calcium scoring have been developed. For instance, I\v{s}gum et al. \cite{ivsgum2010automated} proposed to quantify calcifications in the thoracic aorta by delineating the aorta using a multi-atlas segmentation approach and classifying candidate lesions within this volume with two sequential kNN classifiers. Similarly, Chellamuthu et al. \cite{chellamuthu2017atherosclerotic} used two sequential CNNs to detect whether calcified lesions were present in a number of large arteries in the chest and neck. Simultaneously the location of a bounding box around the lesion was regressed and thereafter a CNN was used to segment the lesion within the bounding box. In both approaches, application of sequential classifiers substantially reduced the amount of false positive lesion detections \cite{ivsgum2010automated,chellamuthu2017atherosclerotic}.

Recently, deep learning based methods were proposed for calcium scoring in low-dose chest CT \cite{lessmann_automatic_2018,lessmann2016deep}. Lessmann et al. \cite{lessmann_automatic_2018} used two subsequent CNNs to detect calcifications in the coronary arteries, aorta and cardiac valves. The first CNN exploited dilated convolutions to identify candidate calcification voxels and label them according to their spatial location, and the second CNN identified true calcified voxels among those detected by the first CNN. Note that the voxels were identified instead of the lesions. This was needed to address low-dose CT where high noise levels don't allow region growing of the voxels above the 130 HU extraction threshold.

The aforementioned methods were typically developed and evaluated in a set of scans with uniform image acquisition characteristics. Due to the supervised nature of these methods, they are sensitive to changes in image acquisition parameters and the imaged population. Consequently, a good performance if often suboptimal in CT scans acquired with different acquisition parameters, which limits the clinical applicability. To extend the applicability, in a large-scale multicenter study including CT scans with diverse acquisition protocols, van Velzen et al. \cite{vanvelzen2020deep} evaluated training strategies for the deep learning method proposed by Lessmann et al. \cite{lessmann_automatic_2018} for calcium scoring in the coronary arteries and the aorta. The performance was evaluated in CSCT, diagnostic chest CT, radiotherapy planning CT and attenuation correction CT acquired with cardiac perfusion PET. Training with a combination of all available types of CT scans led to excellent agreement in risk categorization between manual and automatic scoring, i.e. linearly weighted Cohen's kappa $>$ 0.9 \cite{cohen1960coefficient}. This type of generic method eliminates the need for a specialized network for every type of CT, which is practical for wide application in clinical practice.

\subsection{Calcium scoring in contrast enhanced scans}
\begin{figure}[h]
    \centering
    \includegraphics[width=\textwidth, trim={0cm 0cm 0cm 0cm}, clip]{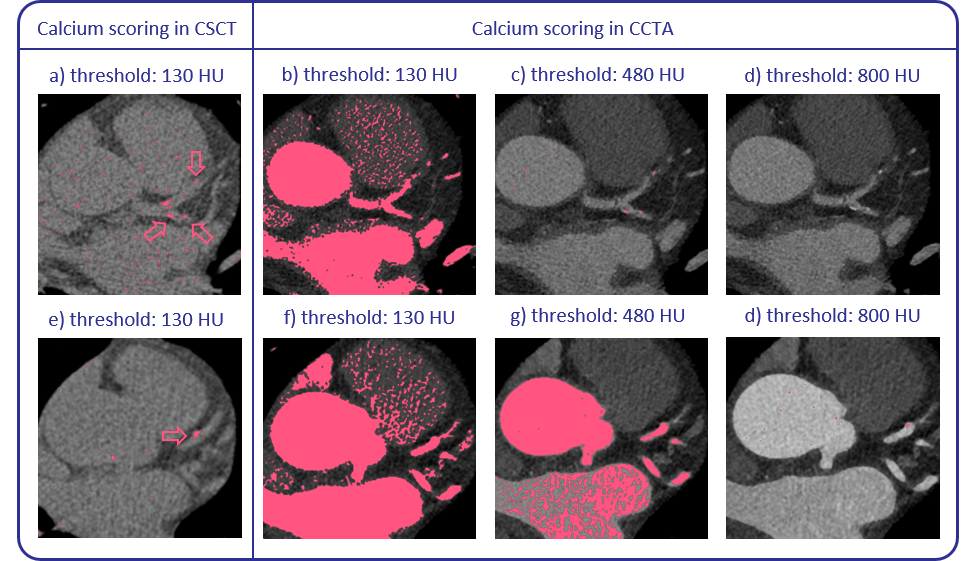}
    \caption{Paired CSCT and CCTA scans for two different patients (top and bottom). In the CSCT scans, arrows indicate lesions detected by using the clinical intensity level threshold of 130 HU. Different thresholds for calcium scoring in CCTA demonstrate inapplicability of a global threshold and thus the need for adjusted approaches.}
    \label{fig:CCTAcac}
\end{figure}

For CCTA acquisition an iodine-containing contrast agent is intravenously administered, which enables visualization of blood vessels in the coronary artery tree. Hence, the current diagnostic purpose of CCTA lies in identification of non-calcified plaque, as well as determination of the presence and quantification of the anatomical significance of coronary artery stenosis. 
Standard calcium scoring is not applicable to CCTA as the contrast agent in the artery typically exceeds the clinical intensity level threshold of 130 HU (Fig. \ref{fig:CCTAcac}). This led to application of higher detection thresholds of for instance 320 HU \cite{otton2012method} or 600 HU \cite{glodny2009method}. However, due to variations between CCTA scans caused by differences in acquisition protocols, scanners and contrast agents, using the same detection threshold for all subjects does often not lead to the most optimal results (Fig. \ref{fig:CCTAcac}). 
To address this, the threshold was derived automatically at standardized anatomical locations, e.g. the ascending aorta \cite{mylonas2014quantifying} or the proximal coronary arteries \cite{pavitt2014deriving}.

Automatic methods for CAC scoring in CCTA typically determine the CAC extraction threshold and localize the coronary arteries (and aorta) by finding their centerlines, thereby allowing detection of CAC lesions in the proximity of the detected centerlines. For example, Teßman et al. \cite{tessmann2011automatic} derived a threshold from the histogram of the localized vessels and applied this threshold to all voxels in the extracted vessel tree. Subsequently, more complex models were developed in which CAC was defined in relation to the local lumen attenuation values \cite{dey2009automated,wesarg2006localizing,ahmed2015automatic}.
Furthermore, Mittal et al. \cite{mittal2010fast} combined machine learning approaches for CAC scoring in CSCT while exploiting the visibility of the coronary arteries in CTA. Namely, the authors trained a random forest classifier based on handcrafted features to identify CAC lesions along an extracted coronary artery centerline.

Later, deep learning approaches were also utilized. Because extraction of the coronary artery centerlines may be challenging especially in presence of pathology like severe stenosis, Wolterink et al. \cite{wolterink_automatic_2016} circumvented this by defining a bounding box around the heart using a deep learning approach \cite{devos2017convnet}. Thereafter, all voxels within the bounding box were classified by an ensamble of four pairs of sequential CNNs, each with either a 2.5D and 3D architecture and varying size of the receptive field. The method showed a high correlation with CSCT reference scores and good agreement of risk categorization \cite{cohen1960coefficient}.  
Zreik et al. \cite{zreik2018recurrent} first extracted coronary artery centerlines and classified the presence of CAC along these centerlines with a combination of a CNN and a recurrent neural network. In this approach CAC was not segmented, but its presence was detected.
A later method by Fischer et al. \cite{fischer2020accuracy} used a similar approach that combined a CNN and a recurrent neural network for CAC detection. While the method showed a high accuracy for CAC lesion detection, their segmentation was not performed and hence, quantification of CAC was not evaluated.

\begin{figure}[t]
    \centering
    \includegraphics[width=\textwidth, trim={1cm 19.5cm 2cm 0.7cm}, clip]{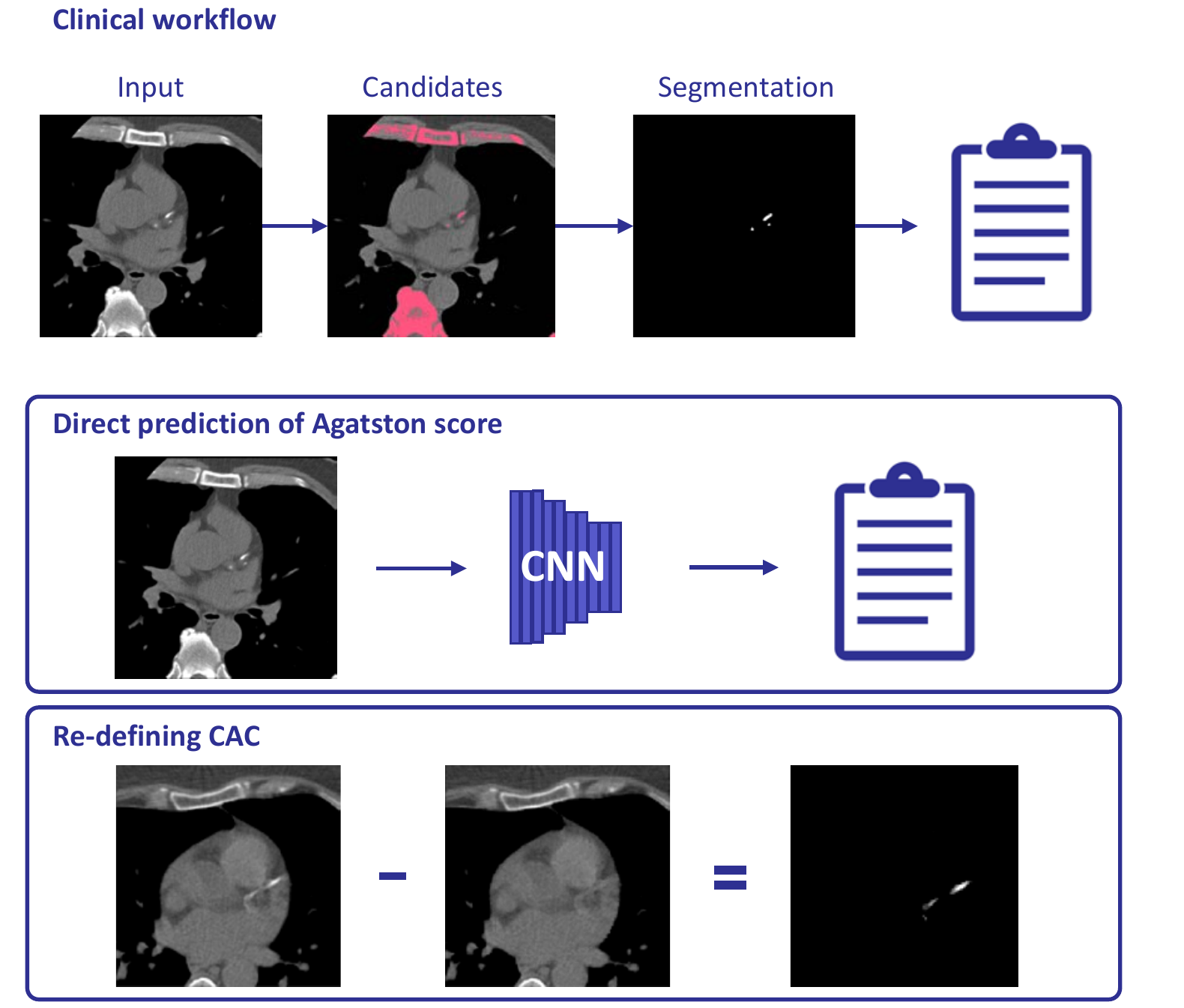}
    \caption{By augmenting the clinical workflow, several automatic methods have been developed that offer new opportunities.}
    \label{fig:advancing}
\end{figure}

\section{Advancing the standard calcium scoring protocol}

While the aforementioned methods employed different methodologies to perform automatic calcium scoring, they all closely followed the clinical workflow using the standard definition of arterial calcification. By adjusting this workflow, several automatic methods have been developed that bring calcium scoring to a new level (Fig. \ref{fig:advancing}). For instance, by circumventing the explicit lesion segmentation and directly predicting an Agatston score, methods can be orders of magnitude faster. Furthermore, revisiting the definition of CAC led to a method that provides more reproducible CAC quantification.

\begin{figure}[h]
    \centering
    \includegraphics[width=\textwidth, trim={1cm 27.5cm 2cm 1cm}, clip]{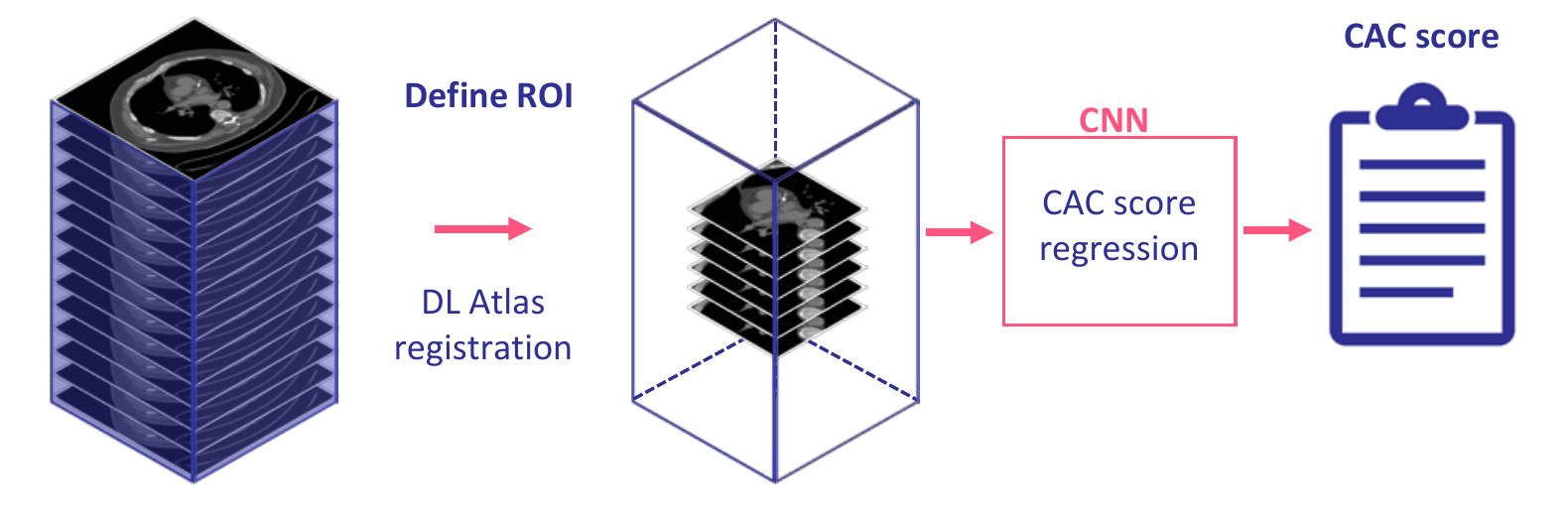}
    \caption{Direct calcium scoring framework by de Vos et al. \cite{vos_direct_2019}: First the region of interest (ROI) containing the heart is determined using deep learning (DL) atlas-registration. Subsequently, the calcium score is regressed with a CNN.}
    \label{fig:direct}
\end{figure}

\subsection{Direct prediction of coronary calcium scores}
While the aforementioned methods aimed at calcium scoring in CSCT and other non-contrast CT showing the heart use different approaches, they all follow the manual calcium scoring workflow: Calcified lesions are first identified and thereafter quantified. Although several of the methods show good performance, this comes at a considerable computational cost. Therefore, CAC scoring methods have been proposed that bypass the lesion segmentation step and, instead, directly regress the Agatston score from the CT image \cite{vos_direct_2019,cano-espinosa_automated_2018}. Like other methods, these methods focused the analysis on the heart by first defining a ROI. Cano-Espinoza et al. \cite{cano-espinosa_automated_2018} defined a bounding box around the heart using an object-detection method. After defining the ROI, the Agatston score was regressed by a 3D CNN that analyzed the heart volume. This means that, instead of a categorical label as was used in classification, a continuous calcium score is predicted. However, in 14\% of the images the heart detection failed and the subsequent analysis could not be performed.
In contrast to Cano-Espinoza et al., de Vos et al. \cite{vos_direct_2019} defined the ROI through registration that aligned the analyzed CT with an atlas image that was constructed from CSCT scans. Conventional image registration is typically computationally expensive and slow, defying the purpose of a direct calcium scoring method. Therefore, a deep learning image registration framework \cite{devos2019deep} was used, which is able to register a 3D image volume in less than a second. Thereafter, the method utilized a 2D CNN architecture to predict an Agatston score in the ROI per image slice. This manner of computation mimicked the clinical calculation of the Agatston score, which is performed in 2D axial image slices (Fig. \ref{fig:direct}). Next to an accuracy in risk categorization that was comparable or better than for other state-of-the-art calcium scoring methods in cardiac and chest CT \cite{wolterink2015automatic,lessmann_automatic_2018}, de Vos et al. \cite{vos_direct_2019} showed that their direct calcium scoring method was hundreds of times faster.

\subsection{Improving reproducibility}
\label{sec:reproducibility}
Although coronary calcium scoring is an established and effective method for CVD risk categorization, its interscan reproducibility is limited. For example, in dedicated CSCT scans that are optimized for CAC scoring, the interscan difference in Agatston scores was reported to range from 15\% to 41\% \cite{detrano2005coronary,mao2001effect,hoffmann2006evidence,van2003coronary}. In chest scans the reproducibility is even more compromised, due to lack of ECG-triggering and the lower image in-plane resolution. In these scans the reported interscan difference is up to 71\%, which led to a difference in risk categorization in 24\% of the subjects \cite{jacobs2010coronary}.

Due to the relatively low in-plane resolution especially in non-cardiac CT scans that visualize the heart, the partial volume effect plays a significant role in CAC quantification. The partial volume effect blurs small calcifications so that they remain below the threshold and are missed during segmentation, causing under-estimation of the amount of CAC. Another cause of limited scoring reproducibility is the fact that non-cardiac CTs, like chest CTs, are acquired without ECG-triggering. In these scans extensive cardiac motion can blur lesions or make them invisible.

Increasing the interscan agreement was addressed by many researchers. The earliest approaches adjusted the intensity level threshold for detection of coronary calcifications, instead of using the fixed 130 HU. Groen et al.\cite{groen2009threshold} proposed to use a lesion specific adaptive threshold that is determined using the maximum intensity of the each CAC lesion in non-ECG-triggered CT \cite{groen2009threshold}. In contrast, Song et al. adapted the theshold in a dedicated CSCT protocol based on the intensity of the background in the vicinity of lesions \cite{song2019improved}. Another approach aimed at dedicated CSCT protocols was proposed by Sauer et al. \cite{saur2009accuratum}, who used a mesh-based algorithm to refine the boundaries of a CAC lesion based on the intensity value profile. However, creating a boundary model for small calcifications or those strongly affected by cardiac motion is very challenging, since typically no intensity plateau is reached because of the partial volume effect, which may hamper performance in CT protocols without ECG-triggering. A different method, aimed at non-ECG-triggered protocols, was proposed by \v{S}prem et al. \cite{vsprem2018coronary}, who built on work by Dehmeshki et al. \cite{dehmeshki2007volumetric}. The authors based their methods on the fact that, due to the limited spatial resolution and cardiac motion, voxels of lesions can contain a mixture of CAC and other tissues. Using an expectation-maximization algorithm they determined the partial calcium content in each voxel of a CAC lesion and its vicinity. Subsequently, the volume of the lesion was corrected by the partial calcium content.

Development and evaluation of partial volume correction methods is often hampered by the fact that the true amount of calcium in patients cannot be measured non-invasively and thus it is not available. Therefore, most partial volume correction methods either used a phantom for development or used cadaver data. This is problematic for training AI methods that require large training sets. Unfortunately, thus far, only a few methods have been shown to translate well to patient data \cite{vsprem2018coronary,dehmeshki2007volumetric}.

\subsection{Revisiting the clinical definition of arterial calcification}

As outlined in Section \ref{sec:reproducibility}, intensity level thresholding contributes to the large interscan reproducibility issues of calcium quantification affected by the partial volume effect and cardiac motion artefacts, especially in non-ECG triggered CT scans.

\begin{figure}[h]
    \centering
    \includegraphics[width=\textwidth, trim={0cm 25cm 0.5cm 1cm}, clip]{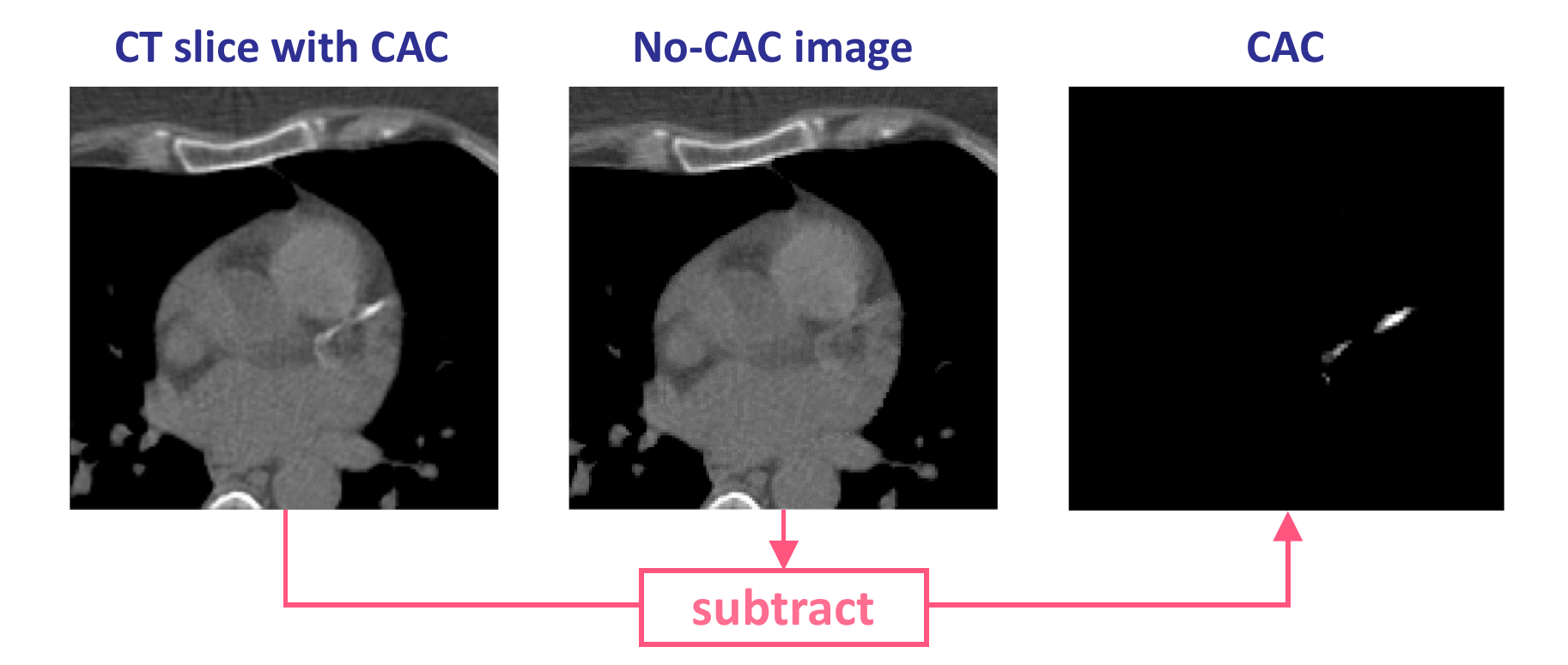}
    \caption{By defining CAC as the difference between an image visibly containing CAC and a corresponding image without CAC, lesions affected by motion and the partial volume effect can be quantified.}
    \label{fig:redefining}
\end{figure}

To address this, an automatic CAC scoring method that abandons the clinically used definition of CAC was proposed \cite{vanvelzen2020coronary}. Instead of defining lesions using the intensity level threshold, the method defined lesions as the difference between an image visibly containing CAC and a corresponding image without CAC (Fig. \ref{fig:redefining}). Because scan pairs of the same patient, where one scan shows CAC and the other one does not, do not exist, image synthesis with a CycleGAN was used to generate images without CAC from images containing CAC. By using labels that indicate whether an image slice contains CAC, image slices can be sorted into two domains: a \textit{containing CAC} domain and a domain \textit{without CAC}. Based on these labels a CycleGAN was trained to translate images from one domain to the other and vice versa. By defining the task in this manner, the use of a threshold for detection and segmentation was avoided. Instead, the model itself decides which voxels contain CAC from a higher level of abstraction. Thereby, the method allows quantification of lesions that are partially or completely below the threshold and implicitly corrects for partial volume effect and motion artefacts. The method led to substantially lower relative interscan difference than standard manual calcium scoring. This may lead to improved and more reliable CVD risk prediction.

\section{Future directions and conclusions}
Recent advances in artificial intelligence research have sparked predictions of a future in which no longer radiologists but advanced computer algorithms read medical images. While this future may still seem distant, AI-based automatic image analysis may much sooner become useful as a supportive tool in the clinic, serving as a second pair of eyes or relieving radiologists and image analysts of tedious tasks.  
Since manual calcium scoring can be tedious and time consuming, a multitude of AI methods have been developed to automate calcium scoring \cite{hampe2019machine,litjens2019state} 
Typically, these AI-based image analysis methods are trained and evaluated on single center studies with high risk for selection biases, and under exclusion of low quality scans. For application in clinical routine, software tools need to be robust to variation in image acquisition parameters and variation in the imaged population. Therefore, ideally large, diverse, well-structured and labeled data sets would be available for the development of new methods and validation of existing techniques. In healthcare, legal barriers often hamper data sharing, and therefore, collecting data for large multi-center evaluation studies is often difficult. This underlines the importance of image analysis challenges for benchmarking the performance of methods. The orCaScore challenge \cite{wolterink2016evaluation} provided an evaluation of several automatic method for calcium scoring in CSCT using multi-center and multi-vendor data. Moreover, a number of methods have proven to be applicable to CTs with different populations or protocols that they were trained on \cite{ivsgum2018automatic,vanvelzen2020deep,gernaat2016automatic}. This versatility and robustness indicates great potential for broad application in clinical practice.

CAC scoring in CSCT is considered a relatively easy task for an expert and the interobserver agreement is very high in these scans\cite{wolterink2016evaluation}. In non-ECG-synchronized CT, where coronary calcium scoring is more challenging due to e.g. noise, low resolution and motion artefacts, the interobserver variability is typically higher. Similar differences over scans are seen in the performance of the automatic methods. While the orCaScore challenge demonstrated excellent performance of automatic methods in CSCT that closely follow expert performance  \cite{wolterink_automatic_2015,vanvelzen2020deep,devos2019direct}, methods applied to other CT scans showing the heart were challenged by image artifacts and low resolution. Nevertheless, also in these cases, the performance of several automatic methods closely followed interobserver agreement \cite{lessmann_automatic_2018,vanvelzen2020deep,devos2019direct}.

Related to the nature of many AI-driven calcium scoring methods, different approaches were used. Early machine learning methods that were aimed at classifying lesions typically exploit lesion-level labels \cite{Isgum2007,kurkure_supervised_2010,qian2010lesion,brunner_toward_2010,shahzad_vessel_2013,wolterink_automatic_2015,isgum_automatic_2012}. Later methods were trained to classify candidate voxels with voxel-level labels \cite{lessmann_automatic_2018,vanvelzen2020deep}. Although both approaches require substantial manual labor for obtaining the reference standard, they provide strong supervision, which have been shown to often lead to the best performance compared to other training strategies. Nevertheless, other training strategies also have been successfully used. Methods that directly regress an Agatston score from an image slice, use Agatston scores as labels \cite{cano-espinosa_automated_2018,devos2019direct}. Because in this approach the labels are defined per image slice instead of per voxel or lesion, the supervision is less strict than with voxel- or lesion labels. However, CAC needs to be manually segmented before a reference Agatston score can be derived, which does not alleviate the workload of defining the reference standard. An explicit advantage of this approach is the gain in speed compared to conventional methods that use segmentation. A different approach, which does not use the clinical CAC definition, uses labels that merely indicate whether CAC is present in an image slice \cite{vanvelzen2020coronary}. This manner of annotation is less labor intensive than CAC segmentation. In this approach, the method is trained to derive which voxels contain CAC from information that only indicates whether CAC is present.

For acceptance and application in clinic, the interpretability of a machine learning method is of key importance. Most calcium scoring methods produce a segmentation map, indicating per voxel whether it is part of a calcified lesion, that was used in further quantification. Although the mechanism for prediction of the segmentation map itself is often hidden in the complex algorithm or network, inspection of the calcification segmentation shows the plausibility and accuracy of the calcium scoring result. On the other hand, direct calcium scoring methods only produce an output Agatston score and, as such, act to a greater extent as a black box. De Vos et al.\cite{devos2019direct} addressed this issue by implementing an optional decision feedback mechanism that shows which parts of the image contributed to the score. Moreover, several types of visualization techniques for CNNs \cite{zeiler2014visualizing,selvaraju2017grad,huo2019coronary} have been proposed to visualize CNN decision making to improve interpretability of CNNs. 

Since the ultimate goal of calcium scoring is to derive a risk of CVD, future research could investigate direct prediction of CVD risk from a CT scan. Hard outcome labels like CVD events or CVD mortality provide a powerful and reliable reference. These labels are defined according to strict clinical protocol and definitions\cite{world1993icd}, which should make them less subject to interobserver variability. Despite the obvious application potential, this area of research is still in its infancy, likely caused by a combination of a the challenging problem and difficulties of obtaining sufficient data. Given the data sets of limited size that were available for this research, thus far developed methods used a two-step approach, where first features describing the image \cite{van2019direct,de2015automatic} were extracted and thereafter, a classifier was trained to classify patients according to the outcomes. However, much like in conventional machine learning, the first stage was not directly coupled to the final prediction goal. Incorporating clinical patient data for multitask learning \cite{guo2020multi} or advanced augmentation techniques \cite{frid2018synthetic} may make end-to-end training with a limited dataset feasible.

In the era of precision medicine it has become increasingly important to quantify disease related parameters to provide a complete overview of the patients physical state. Because it requires a substantial manual effort to quantify all relevant parameters from the plethora of information comprised in a CT scan, automation using AI methods can facilitate and speed up the quantification process. Measuring a broad set of CVD parameters, including calcium scores, may aid medical experts in lifting the practice of personalized medicine to a new level. While the value of careful evaluation should not be forgotten, the incorporation of AI into CVD risk prediction is not a change that clinicians should fear, but rather, one that should be embraced.

\bibliographystyle{ieeetr}
\bibliography{bookchapter}

\end{document}